\documentclass[
    ,final            % use final for the camera ready runs
  ]
  {aipproc}

\layoutstyle{6x9}

\begin{document}

\title{Strange Quark Contribution to the Proton Spin, from Elastic $\vec{e}p$
and $\nu p$ Scattering}

\classification{13.40.Gp,14.20.Dh}

\keywords{strange nucleon form factors}

\author{Stephen Pate}{
  address={Physics Department, New Mexico State University, Las Cruces NM 88003, USA}
}

\begin{abstract}
The strangeness contribution to the vector and axial form
factors of the proton is presented for momentum transfers in the
range $0.45<Q^2<1.0$ GeV$^2$.  The results are obtained via a combined
analysis of forward-scattering parity-violating elastic $\vec{e}p$
asymmetry data from the $G^0$ and HAPPEx experiments at Jefferson Lab, 
and elastic $\nu p$ and $\bar{\nu} p$ scattering data from Experiment 734 at
Brookhaven National Laboratory.  The
combination of the two data sets allows for the simultaneous extraction of
$G_E^s$, $G_M^s$, and $G_A^s$ over a significant range of $Q^2$ for
the very first time.  Determination of the strange axial form factor $G_A^s$
is vital to an understanding of the strange quark contribution to the proton spin.
\end{abstract}

\maketitle

%%%%%%%%%%%%%%%%%%%%%%%%%%%%%%%%%%%%%%%%%%%%
%% MAINMATTER
%%%%%%%%%%%%%%%%%%%%%%%%%%%%%%%%%%%%%%%%%%%%

The strange quark contribution to the proton spin has been a subject of 
investigation ever since the first polarized inclusive deep-inelastic
measurements of the spin-dependent structure function $g_1(x)$ by EMC \cite{Ashman:1989ig}
demonstrated that the Ellis-Jaffe sum rule \cite{Ellis:1973kp,PhysRevD.10.1669.4} 
did not hold true.
Subsequent measurements at CERN and SLAC supported the initial EMC 
measurements, and a global analysis \cite{Filippone:2001ux} of these data
suggested $\Delta s \approx -0.15$.  This analysis carries with it an
unknown theoretical uncertainty because the deep-inelastic data must be
extrapolated to $x=0$ and an assumption of SU(3)-flavor 
symmetry\footnote{See talk by T. Yamanishi on SU(3)-symmetry-breaking effects.}
must be invoked.

In the meantime, the E734 experiment \cite{Ahrens:1987xe} at Brookhaven 
measured the $\nu p$ and $\bar{\nu} p$ elastic scattering cross sections
in the momentum-transfer range $0.45<Q^2<1.05$ GeV$^2$.  These cross sections
are very sensitive to the strange axial form factor of the proton,
$G_A^s(Q^2)$, which is related to the strange quark contribution to the
proton spin:  $G_A^s(Q^2=0)=\Delta s$.  Assuming the strange axial form
factor had the same $Q^2$-dependence as the isovector axial form factor,
E734 also extracted a negative value for $\Delta s$.  However, this
determination was hampered by the large systematic uncertainies in the
cross section measurement, as well as a lack of knowledge of the strange
vector form factors, and no definitive determination of $\Delta s$ was
possible --- this conclusion was confirmed by subsequent reanalyses
of these data \cite{Garvey:1993cg,Alberico:1998qw}.

The HERMES\footnote{See talk by H. Jackson.} experiment \cite{HERMES_deltas}
measured the helicity distribution
of strange quarks, $\Delta s(x)$, using polarized semi-inclusive deep-inelastic
scattering and a leading order ``purity'' analysis, and found 
$\Delta s(x)\approx 0$ in the range $0.03<x<0.3$.  This seems to disagree 
with the analysis of the inclusive deep-inelastic data.  This disagreement 
could be due to a failure of one or more of the assumptions made in the
analysis of the inclusive and/or the semi-inclusive data, or it could be
due to a more exotic physics mechanism such as a ``polarized condensate''
at $x=0$ not observable in deep-inelastic scattering \cite{BassRMP}.

On account of the apparent discrepancy between the two kinds of deep-inelastic
data, another method is needed to shed light on the strange
quark contribution to the proton spin.  Recently \cite{Pate:2003rk} it
has become possible to determine the strange vector and axial form
factors of the proton by combining data from elastic 
parity-violating $\vec{e}p$
scattering experiments at Jefferson Lab with the $\nu p$ and $\bar{\nu} p$
elastic scattering data from E734.  The parity-violating $\vec{e}p$ data place
constraints on the strange vector form factors that were not available
for previous analyses of E734 data.

Several experiments\footnote{See talks by R. Michaels and K. Nakahara.}
have now produced data on forward parity-violating $\vec{e}p$ elastic
scattering \cite{Aniol:2004hp,Maas:2004ta,Maas:2004dh,Armstrong:2005hs,HAPPEx_1H_010,HAPPEx2006}.  
Of most interest here are measurements that lie in the
same $Q^2$ range as the BNL E734 experiment, which are the original
HAPPEx measurement~\cite{Aniol:2004hp} at $Q^2=0.477$~GeV$^2$ and four
points in the recent $G^0$ data~\cite{Armstrong:2005hs}.  These
forward scattering data are most sensitive to $G_E^s$, somewhat less
sensitive to $G_M^s$, and almost completely insensitive to the axial
form factors due to supression by both the weak vector electron charge
$(1-4\sin^2\theta_W)$ and by a kinematic factor that 
approaches 0 at forward angles. 

\begin{figure}[t]
  \includegraphics[height=.48\textheight, bb=150 145 495 520]{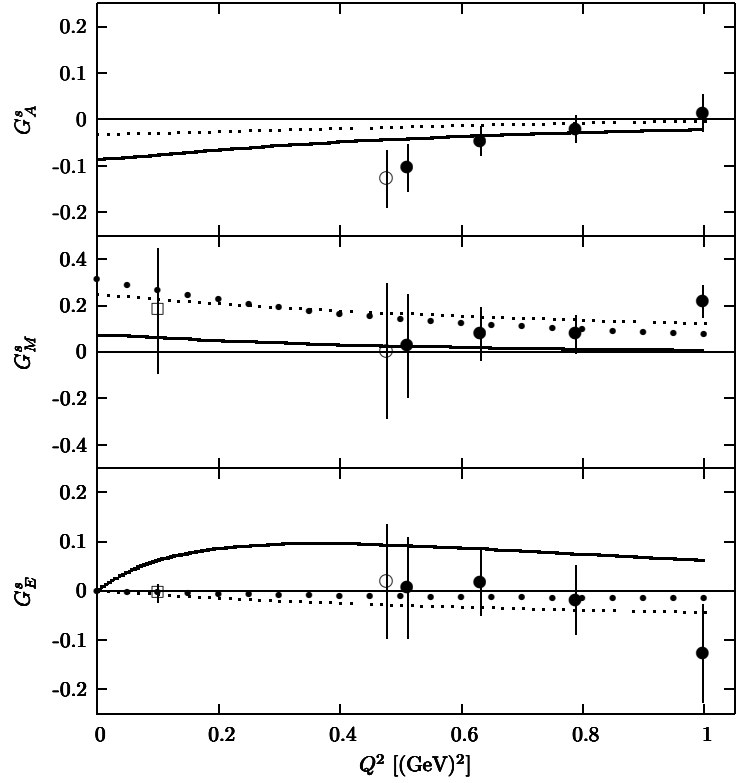}
  \caption{Results of this analysis for the strange vector and axial form factors
of the proton.  Open circles are from a combination of 
HAPPEx and E734 data, while the
closed circles are from a combination of $G^0$ and E734 data.  [Open squares are
from Ref.~\cite{HAPPEx2006} and involve parity-violating $\vec{e}p$ data only.]
The theoretical curves are from Ref.~\cite{Silva:2005fa,Goeke:2006gi,Silva:2005qm} (solid line), 
Ref.~\cite{Riska:2005bh} (small-dotted line), and Ref.~\cite{Bijker:2005pe} 
(big-dotted line).  There is not any calculation of $G_A^s$ from 
Ref.~\cite{Bijker:2005pe}.}
\label{sff_fig}
\end{figure}

The basic technique for combining the $\vec{e}p$, $\nu p$, and $\bar{\nu}p$
data sets has already been
described~\cite{Pate:2003rk} and the details of the present analysis
will be published~\cite{PMP}.  The results are displayed in
Figure~\ref{sff_fig}.  The uncertainties in all three form factors are
dominated by the large uncertainties in the neutrino cross section
data.  Since those data are somewhat insensitive to $G_E^s$ and
$G_M^s$ then the uncertainties in those two form factors are generally
very large.  However the results for the strange axial form factor are of
sufficient precision to give a hint of the $Q^2$-dependence of this
important form factor for the very first time.  There is a strong
indication from this $Q^2$-dependence that $\Delta s < 0$, {\em i.e.}
that the strange quark contribution to the proton spin is negative.
However the data are not of sufficient quality to permit an
extrapolation to $Q^2=0$, so no quantitative evaluation of $\Delta s$
from these data can be made at this time.

It is interesting to compare these results with models that can
calculate a $Q^2$-dependence for these form factors.  Silva, Kim,
Urbano and Goeke~\cite{Silva:2005fa,Goeke:2006gi,Silva:2005qm} have
used the chiral quark soliton model ($\chi$QSM) to calculate
$G^s_{E,M,A}(Q^2)$ in the range $0.0<Q^2<1.0$ GeV$^2$.  The $\chi$QSM
has been very successful in reproducing other properties of light
baryons using only a few parameters which are fixed by other data.  In
Figure~\ref{sff_fig} their calculation is shown as the solid line; it
is seen to be in reasonable agreement with the available data,
although the HAPPEx $G^s_E$ point at $Q^2=0.1$ GeV$^2$ disfavors this
calculation. Riska, An, and
Zou~\cite{Zou:2005xy,An:2005cj,Riska:2005bh} have explored the
stangeness content of the proton by writing all possible $uuds\bar{s}$
configurations and considering their contributions to
$G^s_{E,M,A}(Q^2)$.  They find that a unique $uuds\bar{s}$
configuration, with the $s$ quark in a $P$ state and the
$\bar{s}$ in an $S$ state, gives the best fit to the data for
these form factors; see the small-dotted curves in Figure~\ref{sff_fig}.
Bijker~\cite{Bijker:2005pe} uses a two-component model of
the nucleon to calculate $G^s_{E,M}(Q^2)$; the two components are an
intrinsic three-quark structure and a vector-meson ($\rho$, $\omega$,
and $\phi$) cloud; the strange quark content comes from the meson
cloud component.  The values
of $G^s_{E,M}(Q^2)$ are in good agreement with the data, see the
big-dotted line in Figure~\ref{sff_fig}.  In the near future,
the $G^0$ experiment will provided additional data on $G^s_{E,M}(Q^2)$
at 0.23 abd 0.63 GeV$^2$ which will help to discriminate between the
$\chi$QSM and the models of Bijker and of Riska et al.

To provide a useful determination
of $\Delta s$, better data are needed for both the form factors and the
polarized parton distribution functions.
Two new experiments have been proposed to provide improved neutrino data for
the determination of the strange axial form factor.
FINeSSE~\cite{FINeSSE_FNAL_LOI} proposes to
measure the ratio of the neutral-current
to the charged-current $\nu N$ and $\bar{\nu}N$ processes. A measurement of 
$R_{NC/CC}=\sigma(\nu p\rightarrow\nu p)/\sigma(\nu n\rightarrow\mu^- p)$
and
$\bar{R}_{NC/CC}=\sigma(\bar{\nu}p\rightarrow\bar{\nu}p)/\sigma(\bar{\nu}p\rightarrow\mu^+n)$
combined with the world's data on forward-scattering PV $ep$ data can produce
a dense set of data points for $G_A^s$ in the range $0.25<Q^2<0.75$ GeV$^2$ 
with an uncertainty at each point of about $\pm 0.02$.  
Another experiment with similar physics
goals, called NeuSpin\footnote{See talk by Y. Miyachi.}, is being proposed
for the new J-PARC facility in Japan.
It is also important to extend the semi-inclusive deep-inelastic data
to smaller $x$ and higher $Q^2$ so that the determination of the
polarized strange quark distribution $\Delta s(x)$ can be improved.  A
measurement of this type is envisioned \cite{Stosslein:2000am,Deshpande:2005wd} for the 
proposed electron-ion collider facility.
It is only with these improved data sets that we will be able to arrive
at an understanding of the strange quark contribution to the proton spin.

%%%%%%%%%%%%%%%%%%%%%%%%%%%%%%%%%%%%%%%%%%%%%%%%
%% BACKMATTER
%%%%%%%%%%%%%%%%%%%%%%%%%%%%%%%%%%%%%%%%%%%%%%%%

\begin{theacknowledgments}
The author is grateful to S.D. Bass, H.E. Jackson, and D.O. Riska 
for useful discussions.
This work was supported by the US Department of Energy.
\end{theacknowledgments}

%%%%%%%%%%%%%%%%%%%%%%%%%%%%%%%%%%%%%%%%%%%%%%%%
%% The bibliography can be prepared using the BibTeX program or
%% manually.
%%
%% The code below assumes that BibTeX is used.  If the bibliography is
%% produced without BibTeX comment out the following lines and see the
%% aipguide.pdf for further information.
%%
%% For your convenience a manually coded example is appended
%% after the \end{document}
%%%%%%%%%%%%%%%%%%%%%%%%%%%%%%%%%%%%%%%%%%%%%%%%

%%%%%%%%%%%%%%%%%%%%%%%%%%%%%%%%%%%%%%%%%%%%%%%%
%% You may have to change the BibTeX style below, depending on your
%% setup or preferences.
%%
%%
%% For The AIP proceedings layouts use either
%%%%%%%%%%%%%%%%%%%%%%%%%%%%%%%%%%%%%%%%%%%%

\bibliographystyle{aipproc}   % if natbib is available

\bibliography{my_proc}

%%%%%%%%%%%%%%%%%%%%%%%%%%%%%%%%%%%%%%%%%%%
%% Just a reminder that you may have to run bibtex
%% All of it up to \end{document} can be removed
%% if you don't like the warning.
%%%%%%%%%%%%%%%%%%%%%%%%%%%%%%%%%%%%%%%%%%%
\IfFileExists{\jobname.bbl}{}
 {\typeout{}
  \typeout{******************************************}
  \typeout{** Please run "bibtex \jobname" to optain}
  \typeout{** the bibliography and then re-run LaTeX}
  \typeout{** twice to fix the references!}
  \typeout{******************************************}
  \typeout{}
 }

\end{document}